\PassOptionsToPackage{table}{xcolor}

\documentclass[
  10pt,
  conference,
  letterpaper,
]{IEEEtran}

\IEEEoverridecommandlockouts

\usepackage{amsmath,amssymb,amsfonts}
\usepackage{algorithmic}
\usepackage{graphicx}
\usepackage{textcomp}
\usepackage{xcolor}

\usepackage{cite}

\usepackage{url}

\usepackage{soul}

\usepackage{siunitx}
\DeclareSIUnit{\bps}{bps}

\graphicspath{ {./figs_screenshots} }

\usepackage[caption=false]{subfig}

\usepackage{tikz}
\usetikzlibrary{
	arrows.meta,
	babel,
	backgrounds,
	calc,
	decorations.markings,
	decorations.pathreplacing,
	fit,
	graphs,
	patterns,
	positioning,
	quotes,
	shapes,
	shapes.geometric,
}

\makeatletter
\tikzset{
	fitting node/.style={
			inner sep=0pt,
			fill=none,
			draw=none,
			reset transform,
			fit={(\pgf@pathminx,\pgf@pathminy) (\pgf@pathmaxx,\pgf@pathmaxy)}
		},
	reset transform/.code={\pgftransformreset}
}
\makeatother

\usepackage{pgfplots}
\usepgfplotslibrary{
	patchplots,
	groupplots,
	colorbrewer,
}

\pgfplotsset{compat=newest}
\pgfplotsset{plot coordinates/math parser=false}
\pgfplotsset{every tick label/.style={font=\tiny}}
\pgfplotsset{colormap/Paired}
\pgfplotsset{
	legend image with text/.style={
			legend image code/.code={%
					\node[anchor=center] at (0.3cm,0cm) {#1};
				},
			line width=1pt
		},
}

\usepackage{xspace}

\usepackage[
	acronyms,
	nonumberlist,
	nopostdot,
	nomain,
	nogroupskip,
	acronymlists={hidden},
]{glossaries}
\newacronym{ai}{AI}{Artificial Intelligence}
\newacronym{bs}{BS}{Base Station}
\newacronym{cdf}{CDF}{Cumulative Density Function}
\newacronym{cots}{COTS}{Commercial Off-the-shelf}
\newacronym{cov}{COV}{Coefficient of Variation}
\newacronym{cu}{CU}{Central Unit}
\newacronym{dl}{DL}{downlink}
\newacronym{du}{DU}{Distributed Unit}
\newacronym{ecdf}{ECDF}{Empirical Cumulative Density Function}
\newacronym{embb}{eMBB}{Enhanced Mobile Broadband}
\newacronym{enb}{eNB}{evolved Node Base}
\newacronym{epc}{EPC}{Evolved Packet Core}
\newacronym{es}{ES}{Energy Saving}
\newacronym{fpga}{FPGA}{Field Programmable Gate Array}
\newacronym{int}{INT}{Integral Area}
\newacronym{kpm}{KPM}{Key Performance Metric}
\newacronym{ks}{K-S}{Kolmogorov-Smirnov}
\newacronym{lte}{LTE}{Long Term Evolution}
\newacronym{mgen}{MGEN}{Multi-Generator}
\newacronym{mmtc}{mMTC}{Massive Machine-Type Communications}
\newacronym{mno}{MNO}{Mobile Network Operator}
\newacronym{pdf}{PDF}{Probability Density Function}
\newacronym{prb}{PRB}{Physical Resource Block}
\newacronym{pr}{PF}{Proportional Fair}
\newacronym{ran}{RAN}{Radio Access Network}
\newacronym{ra}{RA}{Resource Allocation}
\newacronym{rbg}{RBG}{Resource Block Group}
\newacronym{rf}{RF}{Radio Frequency}
\newacronym{ric}{RIC}{RAN Intelligent Controller}
\newacronym{nonrtric}{Non-RT RIC}{Non-Real-Time RAN Intelligent Controller}
\newacronym{nearrtric}{Near-RT RIC}{Near-Real-Time RAN Intelligent Controller}
\newacronym{rmssd}{RMSSD}{Root Mean Square of Successive Differences}
\newacronym{rt}{RT}{real-time}
\newacronym{ru}{RU}{Radio Unit}
\newacronym{sdr}{SDR}{Software-Defined Radio}
\newacronym{sla}{SLA}{Service Level Agreement}
\newacronym{smo}{SMO}{Service Management and Orchestration}
\newacronym{srn}{SRN}{Standard Radio Node}
\newacronym{snr}{SNR}{signal-to-noise ratio}
\newacronym{stdev}{SD}{Standard Deviation}
\newacronym{tm}{TM}{Throughput Maximization}
\newacronym{ue}{UE}{User Equipment}
\newacronym{ul}{UL}{Uplink}
\newacronym{urllc}{URLLC}{Ultra Reliable and Low Latency Communications}

\newcommand{\pacifista}{PACIFISTA\xspace}
\newcommand{\oran}{O-RAN\xspace}
\newcommand{\openran}{Open RAN\xspace}

\newcommand{\ran}{\gls{ran}\xspace}

\newcommand{\nonrtric}{\gls{nonrtric}\xspace}
\newcommand{\nearrtric}{\gls{nearrtric}\xspace}

\newcommand{\es}{\gls{es}\xspace}
\newcommand{\tm}{\gls{tm}\xspace}
\newcommand{\severity}{$\sigma$\xspace}

\newcommand{\dks}{D^\textrm{K-S}}
\newcommand{\dint}{D^\textrm{INT}}

\glsdisablehyper

\newlength\fheight
\newlength\fwidth
\usepackage{booktabs}

\usepackage{xfp}

\usepackage[capitalize]{cleveref}
\creflabelformat{equation}{#2#1#3}
\creflabelformat{figure}{#2#1#3}
\creflabelformat{table}{#2#1#3}
\creflabelformat{section}{#2#1#3}

\usepackage{array}
\usepackage{colortbl}
\usepackage{collcell}
\usepackage{hhline}
\usepackage{pgf}
\usepackage{multirow}

\def\colorModel{hsb} %
\newcommand\ColCell[1]{
	\pgfmathparse{#1<0.5?1:0}  %
	\pgfmathsetmacro\compA{(1-#1)/3} %
	\pgfmathsetmacro\compB{0.65} %
	\pgfmathsetmacro\compC{0.9} %

	\edef\x{\noexpand\centering\noexpand\cellcolor[\colorModel]{\compA,\compB,\compC}}\x\relax#1%
}

\newcolumntype{F}{>{\collectcell\ColCell}m{3.0em}<{\endcollectcell}}
\newcolumntype{G}{>{\collectcell\ColCell}m{4.5em}<{\endcollectcell}}

\usepackage[ruled,linesnumbered]{algorithm2e}

\begin{document}

\title{
	Predicting Conflict Impact on Performance\\ in O-RAN
	\thanks{This work was partially supported by OUSD(R\&E) through Army Research Laboratory Cooperative Agreement Number W911NF-24-2-0065. The views and conclusions contained in this document are those of the authors and should not be interpreted as representing the official policies, either expressed or implied, of the Army Research Laboratory or the U.S. Government. The U.S. Government is authorized to reproduce and distribute reprints for Government purposes notwithstanding any copyright notation herein.}
}

\author{
	\IEEEauthorblockN{
		Pietro Brach del Prever$^*$, Niloofar Mohamadi$^*$, Salvatore D'Oro$^*$, Leonardo Bonati$^*$, Michele Polese$^*$,\\ \L ukasz Ku\l acz$^+$, Piotr Jaworski$^+$, Adrian Kliks$^+$, Heiko Lehmann$^\diamond$, Tommaso Melodia$^*$
	}
	\IEEEauthorblockA{$^*$Institute for the Wireless Internet of Things, Northeastern University, Boston, MA, U.S.A.}
	\IEEEauthorblockA{$^+$Poznan University of Technology and Rimedo Labs, Poznan, Poland}
	\IEEEauthorblockA{$^\diamond$Deutsche Telekom AG, T-Labs, Berlin, Germany} \\

}

\maketitle

\begin{picture}(0,0)(-15,-220)
	\put(0,0){
		\put(0,0){\footnotesize
			This paper has been accepted at IEEE International Conference on Computer Communications (INFOCOM 2026).
		}
		\put(0,-10){
			\scriptsize \textcopyright~2026 IEEE. Personal use of this material is permitted. Permission from IEEE must be obtained for all other uses, in any current or future media, including}
		\put(0, -17){
			\scriptsize reprinting/republishing this material for advertising or promotional purposes, creating new collective works, for resale or redistribution to servers or lists,}
		\put(0, -24){
			\scriptsize or reuse of any copyrighted component of this work in other works.}
	}
\end{picture}

\begin{abstract}
	The \oran Alliance promotes the integration of intelligent autonomous agents to control the \ran.
	This improves flexibility, performance, and observability in the \ran, but introduces new challenges, such as the detection and management of conflicts among the intelligent autonomous agents.
	A solution consists of profiling the agents before deployment to gather statistical information about their decision-making behavior, then using the information to estimate the level of conflict among agents with different goals.
	This approach enables determining the occurrence of conflicts among agents, but does not provide information about the impact on \gls{ran} performance, including potential service degradation.
	The problem becomes more complex when agents generate control actions at different timescales, which makes conflict severity hard to predict.
	In this paper, we present a novel approach that fills this gap.
	Our solution leverages the same data used to determine conflict severity but extends its use to predict the impact of such conflicts on \gls{ran} performance based on the frequency at which each agent generates actions, giving more weight to faster applications, which exert control more frequently.
	Via a prototype, we demonstrate that our solution is viable and accurately predicts conflict impact on \ran performance.
\end{abstract}


\glsresetall

\section{Introduction}

The \oran Alliance promotes the integration of \gls{ai} solutions in cellular networks to control and monitor the \ran. Intelligent, autonomous agents are introduced in the \glspl{ric}, which enable a dynamic control of the network at different timescales. In particular, the \textit{rApps}, which run on the \nonrtric, have time loops above $\SI{1}{\second}$, while the \textit{xApps}, which run on the \nearrtric, have time loops between~$\SI{10}{\milli\second}$ and~$\SI{1}{\second}$.

This revolution aims to bring unprecedented flexibility and improved performance, with the ability to observe and control the \gls{ran} at different timescales and reconfigure it based on load, priorities, and business objectives.
However, introducing \gls{ai} into the \gls{ran} also brings challenges, including the coordination of autonomous and independent agents, the management of potential conflicts between them, and the predictability of their interactions.

The \oran Alliance defines such conflicts in \cite{oranwg32025ricarc}. Conflicts are divided into three types: direct, indirect, and implicit. Direct conflicts can be observed directly, and they usually happen when one or more applications target the same control parameter. Indirect conflicts happen when applications do not target the same set of control parameters, but a dependency among such control parameters can be observed (e.g., the radio being turned off indirectly impacts the transmission power). Implicit conflicts, instead, are harder to detect and happen when one application controlling one set of parameters has an effect on another parameter controlled by another application.

Efficiently and accurately detecting, managing, and resolving such conflicts in the \openran context has recently become an active research topic. The solutions proposed so far tackle the problem from various perspectives.

One possible approach is proposed in~\cite{brachdelprever2025pacifista} and consists of profiling autonomous agents in sandboxed environments before deployment, where each xApp (or rApp) is tested individually to characterize its decision-making logic under diverse operational conditions.
However, timescales at which decisions are made by each agent impact the occurrence and severity of conflicts. For example, given two applications, the impact of conflict on \ran performance can significantly vary based on operational conditions. Indeed, conflicts in this case can result in oscillatory behavior if the applications generate control actions at the same frequency (e.g., every $\SI{1}{\second}$), but cause one application's effect to prevail over the other's if one application generates new actions every $\SI{0.1}{\second}$ while the other every $\SI{5}{\second}$.

Being able to predict the impact of conflicts on \gls{ran} performance would enable operators and conflict management systems to evaluate candidate app configurations before deployment, e.g., to determine whether a given set of applications can safely coexist or to select the timescale configuration that minimizes performance degradation. However, this prediction is challenging because it heavily depends on the frequency at which each agent generates actions, which in turn varies with network dynamics.
Empirically estimating this impact generally requires deploying two or more applications at the same time and testing them across different timescale configurations, resulting in combinatorial complexity that is not scalable.
However, a key observation is that, for direct conflicts on shared control parameters, the \gls{ran} state at any given time is determined by the most recent control action.
The resulting parameter distribution can therefore be approximated as a frequency-weighted combination of each application's individual action distribution, suggesting that independently collected application profiles can be statistically combined to estimate conflict impact without simultaneous execution, provided that each application's decision-making logic is not significantly altered by the actions of others.

In this paper, we leverage this observation to estimate a-priori the impact of conflicts on \gls{ran} performance across different timescale configurations using only statistical data collected independently for each application.
We specify the \oran integration and workflow pipeline of this system and validate our approach through experiments with two conflicting xApps.
Results demonstrate that our method can effectively forecast the performance impact of timescale-dependent conflicts in unseen configurations using solely pre-collected statistical data.

The remaining of this paper is organized as follows.
Section~\ref{sec:related_work} surveys related works on conflict mitigation.
Section~\ref{sec:prediction_system} introduces the algorithm used for the prediction of the \ran performance.
Section~\ref{sec:experimental_setup} presents the experimental setup that we used for our evaluation, including the testbed setup (Section~\ref{sec:testbed_setup}),
the xApps used in our experiments (Section~\ref{sec:xapps_description}), and their characterization and the conflicts between them (Section~\ref{sec:xapsp_characterization}).
Finally, Section~\ref{sec:results} discusses results of our experiments, while Section~\ref{sec:conclusions} concludes our work.

\section{Related Work}\label{sec:related_work}

\subsection{Conflict Mitigation in \oran}

In \cite{wadud2023conflict} the authors propose an architecture with two controllers for detection and mitigation, and approach the problem with cooperative bargain game theory. Via experimental results, they demonstrate the effectiveness of finding the optimal configuration for the RAN parameter under conflict, achieving successful mitigation. However, the authors only tested this system in a theoretical scenario and do not assure applicability in the real world.

In \cite{adamczyk2023conflict} and \cite{sultana2025experimental}, a conflict resolution and management framework is proposed. The former focuses on evaluating the framework via simulations, while the latter focuses on its experimental evaluation with \gls{cots} devices (e.g., smartphones). Results in both cases show that the use of conflict management frameworks can improve performance metrics in both cases. The framework presented and tested in these two works is reactive rather than proactive, focusing on online mitigation, and only treats direct conflicts.

\cite{brachdelprever2025pacifista} presents \pacifista, a framework to detect, characterize, and mitigate conflicts pre-deployment. \pacifista performs statistical analysis on data collected in a testing environment for each application to estimate conflict occurrence and severity between applications, notifying operators about potential performance deterioration due to conflicts. \pacifista has a proactive approach to conflict mitigation that is based on the behavior of the xApps in the \ran, but it cannot predict the expected \ran performance, thus lacking the possibility to verify that the prediction correspond to the actual \ran performance, and review the \glspl{ric} configuration if not. \pacifista is described in further detail in Section~\ref{sec:pacifista_primer}.

In \cite{giannopoulos2025comix}, the authors propose the Conflict Management scheme for Multi-Channel Power Control in \oran xApps called COMIX. This approach focuses on the detection and resolution of conflicts between xApps, where one application optimizes data rate in the network, whereas the other energy efficiency.
The authors propose the use of a digital twin for policy validation before any decision made by the conflict mitigation module is applied to the live network.

In \cite{sharma2025towards}, the authors employ explainable machine learning and causal inference to evaluate xApp conflicts. Their approach characterizes xApps by identifying the control parameters that influence specific \glspl{kpm}, then uses causal graphs to estimate the average treatment effect of conflicting control actions. This statistical pre-deployment analysis shares conceptual similarities, and drawbacks, with the profiling approach adopted in~\cite{brachdelprever2025pacifista}.

In \cite{erdol2025xapp}, the authors propose a knowledge distillation approach to merge multiple \gls{ai}-based xApps into a unified model that inherently avoids conflicts. By distilling the behavioral patterns of individual xApps, the resulting model captures the combined decision-making logic without requiring runtime conflict resolution. This method also focuses on proactive conflict mitigation and does not allow to verify if the behavior of the distilled xApp is as intended.

Overall, existing systems focus either on reactive conflict mitigation (online, post-deployment) or proactive conflict mitigation (offline, pre-deployment), and systems with a proactive approach lack mechanisms to verify whether conflicts among coexisting xApps result in tolerable performance degradation. Most importantly, none of these systems link proactive analysis to be used to inform online conflict mitigation, which is the key contribution we present in this paper.

\subsection{Primer on \pacifista}\label{sec:pacifista_primer}

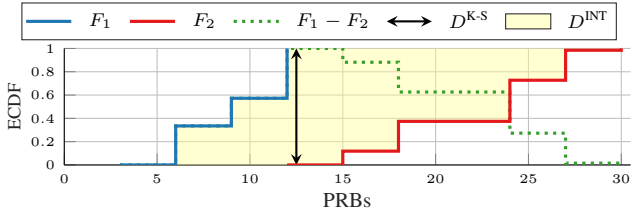
\begin{figure}[htbp]
	\centering
	\setlength{\fwidth}{0.85\columnwidth}
	\begin{tikzpicture}

	\begin{axis}[
			width=\fwidth,
			height=0.35\fheight,
			at={(0\linewidth,0\linewidth)},
			scale only axis,
			xmin=0,
			xmax=30.5,
			xlabel style={font=\footnotesize\color{white!15!black}},
			xlabel={PRBs},
			ymin=0,
			ymax=1,
			ylabel style={font=\scriptsize\color{white!15!black}},
			ylabel={ECDF},
			axis background/.style={fill=white},
			axis x line*=bottom,
			axis y line*=left,
			xmajorgrids,
			ymajorgrids,
			legend style={
                inner sep=1pt,
                legend cell align=left,
                column sep=4pt,
                align=left,
                draw=white!15!black,
                font=\scriptsize,
				at={(0.45,1.07)}, anchor=south,
				},
			legend columns=5,
			reverse legend,
			ylabel shift=-3pt,
			xlabel shift=-3pt,
			clip mode=individual  %
		]

		\addplot[
			const plot,
			fill=Paired-K,
			fill opacity=0.45,
			draw=none,
			area legend,
		]
		table[row sep=crcr] {
				x	y\\
				3	0\\
				6	0.335\\
				9	0.5724\\
				12	1\\
				15	1\\
				18	1\\
				21	1\\
				24	1\\
				27	1\\
				30	1\\
				30	0.9846\\
				27	0.7272\\
				24	0.3741\\
				21	0.3741\\
				18	0.1184\\
				15	0\\
				12	0\\
			}--cycle;
		\addlegendentry{$\dint$}

		\addplot [color=black, thick,
		{stealth[length=2mm,inset=0pt]}-{stealth[length=2mm,inset=0pt]}]
		table[row sep=crcr] {
				12.5 0.0\\
				12.5 1.0\\
			};
		\addlegendentry{$\dks$}

		\addplot [
			const plot,
			color=Paired-D,
			very thick,
			dotted,
		]
		table[row sep=crcr]{
				3	0\\
				6	0.335\\
				9	0.5724\\
				12	1\\
				15	0.8816\\
				18	0.6259\\
				21	0.6259\\
				24	0.2728\\
				27	0.0154\\
				30	0\\
			};
		\addlegendentry{$F_1-F_2$}

		\addplot [
			const plot,
			color=Paired-F,
			very thick,
		]
		table[col sep=tab] {figurestikz/data/ecdf_profiles/sliceprb-4.tsv};
		\addlegendentry{$F_2$}

		\addplot [
			const plot,
			color=Paired-B,
			very thick,
		]
		table[col sep=tab] {figurestikz/data/ecdf_profiles/sliceprb-3.tsv};
		\addlegendentry{$F_1$}

	\end{axis}

\end{tikzpicture}
	\caption{Graphical representation of the different distance metrics.}
	\label{fig:distances_example}
\end{figure}

To evaluate conflicts among applications and their severity, we adopt the \pacifista framework described in~\cite{brachdelprever2025pacifista}.
\pacifista runs in the \gls{smo} component and is composed of four different modules for profiling, detecting, evaluating, and mitigating conflicts.
This framework allows operators to profile individual applications and later determine how well they would coexist with other applications by evaluating the differences between the control actions each of them would take. Profiling data is stored as a time series of \ran variables taken at regular time intervals.
A profiling module takes the collected variables and generates a statistical summary---based on the application \glspl{ecdf} and called \textit{profile}---that includes the distribution of actions taken by the application under certain operational conditions of interest. An illustrative example of profiles of two applications allocating \glspl{prb} according to two different policies is shown in Figure~\ref{fig:distances_example} (blue and red solid lines).

While the profiling module is run only once for each new application in the application catalog, the detection, evaluation, and mitigation modules are executed every time a new application is deployed by the operator.
These modules generate a pair of artifacts from the application profiles they get as input: (i) a \textit{conflict report}, which contains a conflict-level severity index $\sigma^{\mathrm{K}} \in [0, 1]$ (the higher the value, the higher the conflict) for each pair of applications; and (ii) a \textit{conflict-aware instantiation policy}, which takes into account the set of applications that the operator wants to deploy and returns the subset of applications that can be deployed concurrently with an acceptable level of conflict.

Conflict levels are computed from the applications profiles and distances among their \glspl{ecdf}, where different statistical functions return different distances, e.g., the $\dks$ distance based on the \gls{ks} statistical test, and $\dint$ based on the \gls{int} between the two \glspl{ecdf}. Figure~\ref{fig:distances_example} also shows both these distances.
\gls{ks} distance consists of the maximum vertical difference between two \glspl{ecdf}, while the \gls{int} distance consists of the area comprised between the two \glspl{ecdf} divided by the span of the support vector.
A high \gls{ks} distance value for a specific variable indicates the presence of a conflict for that variable for the considered pair of applications, while a low value means that the conflict is absent or minimal. However, the \gls{ks} distance usually presents very high values close to $1$ in the presence of a conflict. Therefore, the \gls{int} distance is used to evaluate the conflict level and compute the severity index \severity.

Albeit providing an indication of conflicts between applications to be deployed, \pacifista focuses on estimating how actions taken by conflicting applications would differ but does not predict how \ran performance would vary as a result of such differences, which is instead the focus of this work.

\section{Prediction System}\label{sec:prediction_system}

\begin{figure}[htbp!]
	\centering
	\includegraphics[width=\linewidth]{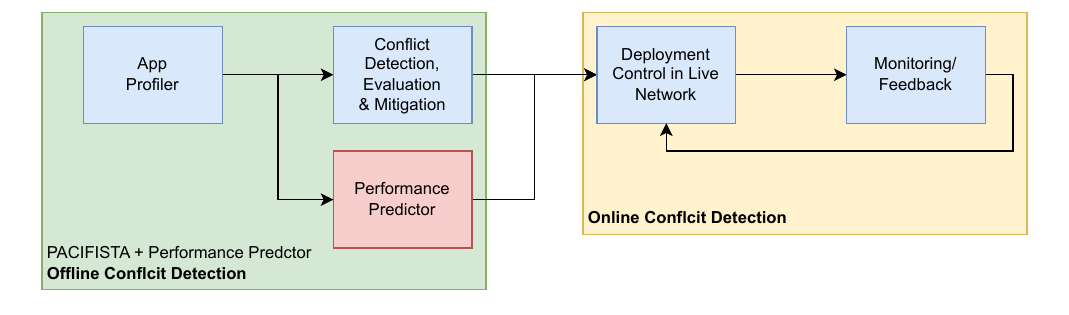}
	\caption{Prediction system placement and workflow within the \oran architecture and relation with \pacifista.}
	\label{fig:system_architecture}
\end{figure}

Figure~\ref{fig:system_architecture} illustrates our performance prediction system and its integration within the conflict mitigation lifecycle. The blue blocks show the end-to-end workflow, from pre-deployment conflict analysis through offline profiling to post-deployment monitoring and control in live networks. Since applications directly affect operational network functions, unresolved conflicts can severely impact network performance and stability, making continuous evaluation critical.

The predictor module (red block) runs in the \gls{smo} and takes as input the application profiles from \pacifista to predict the \gls{cdf}, or \textit{predicted profile}, for each variable (control parameter or \gls{kpm}). Building on \pacifista's severity index \severity, the prediction captures how conflicting objectives at different timescales interact and impact \ran \glspl{kpm} when multiple applications run concurrently. Central to this approach is the control action frequency: faster applications exert influence more often and thus have greater impact on the \ran. Unlike conflicts identifiable offline, certain interactions only emerge in live environments, which the predictor is designed to capture.

As network conditions evolve, the predictor operates both offline---providing deployment insights to operators---and online, where it continuously monitors conditions, re-evaluates predictions, and adapts based on real-time status to enable informed conflict mitigation decisions.

The core of the predictor is Algorithm~\ref{alg:weighted_average}, which computes the predicted profile by taking application profiles and their configured message period (i.e., interval at which applications generate new control actions) as input.
The algorithm evaluates severity for each variable individually---either control parameter or \gls{kpm}---such as \glspl{prb} assignment or Downlink Throughput.

For each profile $P_i$ of the $i$-th application, the \gls{ecdf} $(F_i, x_i)$ is computed. $x_i$ is the vector of sorted observed values from the profile, and $F_i$ is the vector of empirical cumulative probabilities at each point of $x_i$.
To account for the frequency at which each application generates control actions, weights $w_i$ are assigned to each \gls{ecdf} based on the message period $\tau_i$.
Applications with shorter periods (i.e., more frequent control messages) receive higher weights.
The weight for application $i$ is computed as $w_i = ({1/\tau_i})/({\sum_{j=1}^{n} 1/\tau_j})$.

To combine the \glspl{ecdf} from different applications, a common support vector $\mathbf{x}$ is constructed by taking the union of all individual support vectors: $\mathbf{x} = \bigcup_{i=1}^{n} x_i$.
Each \gls{ecdf} $F_i$ is then step-interpolated onto $\mathbf{x}$ to produce $\tilde{F}_i$, where $\tilde{F}_i(x) = F_i(\max\{x' \in x_i : x' \leq x\})$.
The final weighted \gls{ecdf} is obtained as $F_w(x) = \sum_{i=1}^{n} w_i \cdot \tilde{F}_i(x)$, which combines the individual distributions according to their relative frequencies.

\begin{algorithm}[htbp]
	\caption{Weighted \gls{ecdf} Average}
	\label{alg:weighted_average}
	\KwIn{$\mathcal{P} = \{P_1, P_2, \ldots, P_n\}$: profiles}
	\KwIn{$\mathbf{\tau} = [\tau_1, \tau_2, \ldots, \tau_n]$: message periods}
	\KwOut{$F_w(x)$: weighted ECDF at common support points $\mathbf{x}$}
	\For{$i = 1$ \KwTo $n$}{
		$(F_i, \mathbf{x}_i) \gets \mathrm{ECDF}(P_i)$\;
	}
	$w_i \gets \dfrac{1/\tau_i}{\sum_{j=1}^{n} 1/\tau_j}$ \quad for $i = 1, \ldots, n$\;
	$\mathbf{x} \gets \bigcup_{i=1}^{n} \mathbf{x}_i$\;
	$F_w(x) \gets 0$ \quad for all $x \in \mathbf{x}$\;
	\For{$i = 1$ \KwTo $n$}{
		$\tilde{F}_i \gets \mathrm{StepInterpolate}(F_i, \mathbf{x}_i, \mathbf{x})$\;
		$F_w \gets F_w + w_i \cdot \tilde{F}_i$\;
	}
	\Return{$F_w$}
\end{algorithm}

Algorithm~\ref{alg:weighted_average} assumes that application actions are enforced at all times with weights $w_i$ inversely proportional to their message periodicity $\tau_i$.
However, it is worth mentioning that there are edge cases when applications send control messages at the same frequency with a small time offset $\epsilon$, which may result in one application dominating the other.
If two synchronized applications differ by $\epsilon$, the first application's action is applied in the interval $[\tau_i,\tau_i+\epsilon)$, while the second application's action is applied in $[\tau_i+\epsilon,\tau_{i+1})$.
As $\epsilon \rightarrow 0$, the first application's actions are effectively ignored.

To address this timing effect and improve prediction accuracy, the weights $w_i$ can be adjusted to reflect the actual time each action remains active, yielding an effective periodicity $\tau_i^{\text{eff}}$.
For instance, if both applications generate actions every \SI{1}{\second}, but one remains active for \SI{800}{\milli\second} and the other for \SI{200}{\milli\second}, their effective periodicity becomes $\tau_1^{\text{eff}} = \SI{0.8}{\second}$ and $\tau_2^{\text{eff}} = \SI{0.2}{\second}$, respectively.
This refinement captures the real-time impact of each application by weighting not only how frequently actions are generated but also their actual duration in the system.
In practice, this adjustment can be achieved
by collecting sufficient batches of samples to capture variability in $\epsilon$ values, ensuring that the predictor accurately reflects the temporal dynamics of concurrent applications.

\section{Experimental Setup}\label{sec:experimental_setup}

In this section, we describe the experimental setup used for the evaluation of our proposed solution. Section~\ref{sec:testbed_setup} describes our testbed setup, Section~\ref{sec:xapps_description} the xApps we use, and Section~\ref{sec:xapsp_characterization} the characterization of the xApps and their level of conflict.

\subsection{Testbed Setup}\label{sec:testbed_setup}

We leverage the Colosseum testbed~\cite{polese2024colosseum}, an Open \gls{ran} digital twin with hardware-in-the-loop, and the OpenRAN Gym framework~\cite{bonati2023openran}. We consider a topology similar to~\cite{brachdelprever2025pacifista} with 3 \glspl{ue} and 1 \gls{bs}, and a channel bandwidth of \SI{10}{\mega\hertz} with 50 \glspl{prb} grouped into 17 \glspl{rbg}.
We consider a Colosseum scenario that emulates the urban environment in the center of the city of Rome, Italy, where the \glspl{ue} move at moderate speed within a close range from the \gls{bs}.
We generate \gls{dl} traffic for the \glspl{ue} using the \texttt{iperf3} tool. The first \gls{ue} (belonging to the \gls{embb} slice) has a target \gls{dl} bitrate of \SI{2}{\mega\bps}, the second one (\gls{mmtc} slice) of \SI{1}{\mega\bps}, and the third one (\gls{urllc} slice) of \SI{0.5}{\mega\bps}.

We leverage \pacifista's statistical profiling for xApp characterization and conflict evaluation, where the severity index \severity is computed based on the \gls{int} distances among application \glspl{ecdf} (see Section~\ref{sec:pacifista_primer}).
Specifically, \severity is computed as the mean \gls{int} distance of the throughput and buffer occupancy \glspl{kpm}, chosen for their correlation with all other \glspl{kpm} relevant to performance assessment.

\subsection{Considered xApps}\label{sec:xapps_description}

For our analysis, we consider two xApps designed to generate direct conflicts by tuning the amount of \glspl{prb} allocated to the slices of the \gls{bs}.
The first xApp, \es, minimizes the energy used for data transmission at the \gls{bs} by reducing \glspl{prb} assigned to each slice in downlink.
The second, \tm, maximizes throughput by increasing \glspl{prb} to requesting users.
These opposing objectives result in a direct conflict on \gls{prb} allocation that serves as a stress test for the proposed prediction method.
Both xApps incrementally increase/decrease \glspl{prb} by $m$ units ($m=3$ in our experiments).
The frequency $f_i$ at which each xApp generates control actions can be configured to evaluate how action generation rates impact conflict severity and \gls{ran} performance.
Conflicts are characterized using the statistical features of assigned \glspl{prb}, buffer occupancy, and throughput.

\subsection{xApp Characterization and Conflicts}\label{sec:xapsp_characterization}

In this section, we analyze collected data for each application on the Colosseum testbed and specifically in the setup described in Section~\ref{sec:testbed_setup}.
Data is collected in multiple batches, with each batch lasting at least $10$ minutes of live experiment on Colosseum.
Before providing results of the \gls{kpm} prediction module, in this section, we use the collected data to describe the statistical profiles of the two xApps used in our experiments.

\begin{figure}[t!]
	\centering
	\subfloat[\glspl{prb} assigned.]{
		\centering
		\begin{tikzpicture}

	\begin{axis}[%
			width=\fwidth,
			height=0.3\fheight,
			at={(0\fwidth,0\fheight)},
			scale only axis,
			xmin=0,
			xmax=30,
			xlabel style={font=\footnotesize\color{white!15!black}},
			xlabel={PRBs},
			ymin=0,
			ymax=1,
			ylabel style={font=\footnotesize\color{white!15!black}},
			ylabel={ECDF},
			axis background/.style={fill=white},
			axis x line*=bottom,
			axis y line*=left,
			xmajorgrids,
			ymajorgrids,
			legend style={
					inner sep=1pt,
					legend cell align=left,
					align=left,
					draw=white!15!black,
					font=\scriptsize,
					at={(0.5,1.07)}, anchor=south,
				},
			legend columns=4,
			enlargelimits=false,
			clip mode=individual,  %
		]

		\addplot[
			const plot,
			color=Paired-B,
			very thick,
		]
		table[col sep=tab] {figurestikz/data/ecdf_profiles/sliceprb-3.tsv};
		\addlegendentry{ES}

		\addplot[
			const plot,
			color=Paired-F,
			very thick,
		]
		table[col sep=tab] {figurestikz/data/ecdf_profiles/sliceprb-4.tsv};
		\addlegendentry{TM}

    \end{axis}

\end{tikzpicture}%
		\label{fig:ecdf_profiles:sliceprb}
	}

	\subfloat[\gls{dl} buffer occupancy.]{
		\centering
		\begin{tikzpicture}

	\begin{axis}[%
			width=\fwidth,
			height=0.3\fheight,
			at={(0\fwidth,-0.33\fheight)},
			scale only axis,
			xmin=0,
			xmax=2e+05,
			xlabel style={font=\footnotesize\color{white!15!black}},
			xlabel={Buffer Size [Byte]},
			ymin=0,
			ymax=1,
			ylabel style={font=\footnotesize\color{white!15!black}},
			ylabel={ECDF},
			axis background/.style={fill=white},
			axis x line*=bottom,
			axis y line*=left,
			xmajorgrids,
			ymajorgrids,
			enlargelimits=false,
			clip mode=individual,  %
		]

		\addplot[
			const plot,
			color=Paired-B,
			very thick,
		]
		table[col sep=tab] {figurestikz/data/ecdf_profiles/dlbufferbytes-3.tsv};

		\addplot[
			const plot,
			color=Paired-F,
			very thick,
		]
		table[col sep=tab] {figurestikz/data/ecdf_profiles/dlbufferbytes-4.tsv};

	\end{axis}

\end{tikzpicture}%
		\label{fig:ecdf_profiles:dlbufferbytes}
	}

	\subfloat[\gls{dl} throughput.]{
		\centering
		\begin{tikzpicture}

	\begin{axis}[%
			width=\fwidth,
			height=0.3\fheight,
			at={(0\fwidth,0\fheight)},
			scale only axis,
			xmin=0,
			xmax=14,
			xlabel style={font=\footnotesize\color{white!15!black}},
			xlabel={Throughput [Mbps]},
			ymin=0,
			ymax=1,
			ylabel style={font=\footnotesize\color{white!15!black}},
			ylabel={ECDF},
			axis background/.style={fill=white},
			axis x line*=bottom,
			axis y line*=left,
			xmajorgrids,
			ymajorgrids,
			enlargelimits=false,
			clip mode=individual,  %
		]

		\addplot[
			const plot,
			color=Paired-B,
			very thick,
		]
		table[col sep=tab] {figurestikz/data/ecdf_profiles/txbrateDownlinkMbps-3.tsv};

		\addplot[
			const plot,
			color=Paired-F,
			very thick,
		]
		table[col sep=tab] {figurestikz/data/ecdf_profiles/txbrateDownlinkMbps-4.tsv};

	\end{axis}

\end{tikzpicture}%
		\label{fig:ecdf_profiles:txbrateDownlinkMbps}
	}

	\caption{Profiles of xApps run individually in scenarios BASE and ROME: \gls{ue} with \SI{2}{\mega\bps} traffic.}
	\label{fig:ecdf_profiles}
\end{figure}
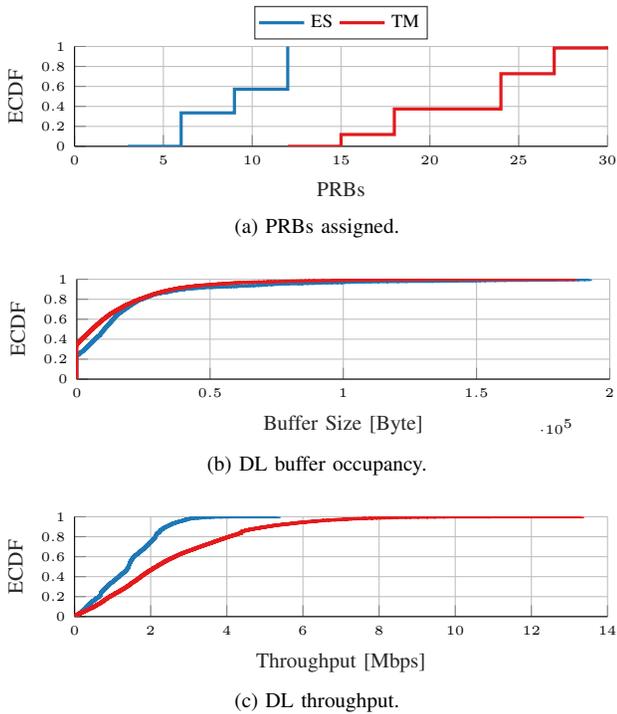

Figure~\ref{fig:ecdf_profiles} shows the profiles of the two xApps with respect to a control parameter (slice \gls{prb} allocation in Figure~\ref{fig:ecdf_profiles:sliceprb}) and two \glspl{kpm} (buffer occupancy in Figure~\ref{fig:ecdf_profiles:dlbufferbytes}, and throughput in Figure~\ref{fig:ecdf_profiles:txbrateDownlinkMbps}).
The differences in the distributions (especially with respect to the control action) suggest that the two applications differ in the way they allocate resources. As explained in Section~\ref{sec:xapps_description}, the \es xApp allocates a small amount of resources to save energy, while the \tm xApp allocates more \glspl{prb} to improve throughput.

Table~\ref{tab:combined_distances_apps}
reports the \gls{ks} and \gls{int} distances (Section~\ref{sec:pacifista_primer}) and the severity indexes  between the two xApps.
As expected, \gls{prb} allocation between the two xApps is very different, with high values of both \gls{ks} and \gls{int} distance. On the other hand, the conflict values of the two \glspl{kpm} considered, i.e., buffer occupancy and throughput, are not as high, as shown in Table~\ref{tab:combined_distances_apps}.
Although conflict values are high for the control parameter (i.e., the number of \glspl{prb} allocated to each slice), this does not necessarily mean that \glspl{kpm} are significantly different.
In our setup, the combination of \gls{rf} and traffic scenarios attenuates how conflicts propagate from control parameters to \glspl{kpm}: high conflicts in \gls{prb} allocation do not translate to proportionally high conflicts in buffer occupancy and throughput.

\begin{table}[htbp]
	\centering
	\caption{\gls{ks} and \gls{int} distances, and severity index (\gls{embb} slice)}
	\label{tab:combined_distances_apps}
	\includegraphics[width=1.0\columnwidth]{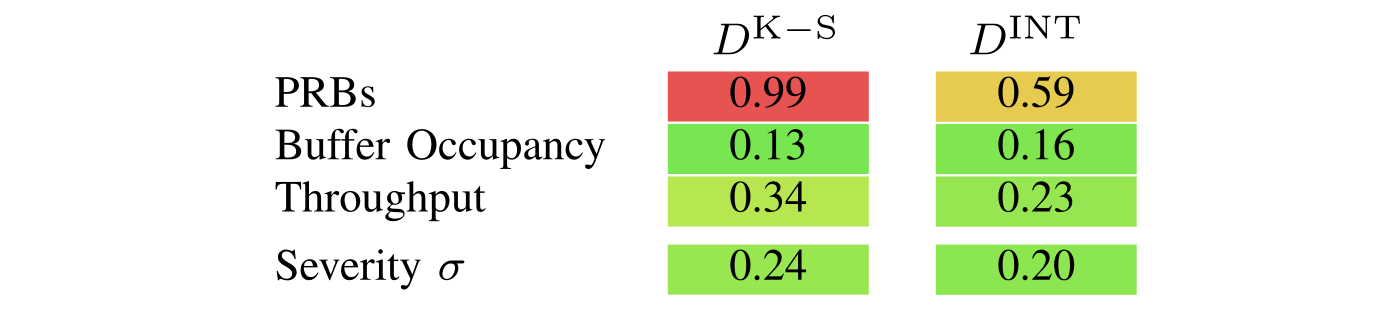}
\end{table}

\begin{figure}[t!]
	\centering
	\subfloat[\glspl{prb} assigned.]{
		\centering
		\begin{tikzpicture}

	\begin{axis}[%
			width=\fwidth,
			height=0.3\fheight,
			at={(0\fwidth,0\fheight)},
			scale only axis,
			xmin=0,
			xmax=33,
			xlabel style={font=\footnotesize\color{white!15!black}},
			xlabel={PRBs},
			ymin=0,
			ymax=1,
			ylabel style={font=\footnotesize\color{white!15!black}},
			ylabel={ECDF},
			axis background/.style={fill=white},
			axis x line*=bottom,
			axis y line*=left,
			xmajorgrids,
			ymajorgrids,
			legend style={
					inner sep=1pt,
					legend cell align=left,
					align=left,
					draw=white!15!black,
					font=\scriptsize,
					at={(0.5,1.07)}, anchor=south,
				},
			legend columns=4,
			enlargelimits=false,
			clip mode=individual,  %
		]
		\addplot[
			const plot,
			color=Paired-B,
			very thick,
		]
		table[col sep=tab] {figurestikz/data/ecdf_concurrentsame/sliceprb-1.tsv};

		\addplot[
			const plot,
			color=Paired-F,
			very thick,
		]
		table[col sep=tab] {figurestikz/data/ecdf_concurrentsame/sliceprb-2.tsv};

		\addplot[
			const plot,
			color=Paired-D,
			very thick,
		]
		table[col sep=tab] {figurestikz/data/ecdf_concurrentsame/sliceprb-3.tsv};

		\addplot[
			const plot,
			color=Paired-C,
			very thick,
			dashed,
		]
		table[col sep=tab] {figurestikz/data/ecdf_concurrentsame/sliceprb-4.tsv};

		\addlegendentry{ES}
		\addlegendentry{TM}
		\addlegendentry{ES1-TM1}
		\addlegendentry{ES1-TM1 Prediction}

	\end{axis}
\end{tikzpicture}%
	}

	\subfloat[\gls{dl} buffer occupancy.]{
		\centering
		\begin{tikzpicture}

	\begin{axis}[%
			width=\fwidth,
			height=0.3\fheight,
			at={(0\fwidth,-0.33\fheight)},
			scale only axis,
			xmin=0,
			xmax=2e+05,
			xlabel style={font=\footnotesize\color{white!15!black}},
			xlabel={Buffer Size [Byte]},
			ymin=0,
			ymax=1,
			ylabel style={font=\footnotesize\color{white!15!black}},
			ylabel={ECDF},
			axis background/.style={fill=white},
			axis x line*=bottom,
			axis y line*=left,
			xmajorgrids,
			ymajorgrids,
			legend style={
					inner sep=1pt,
					legend cell align=left,
					align=left,
					draw=white!15!black,
					font=\scriptsize,
					at={(0.95,0.05)}, anchor=south east,
				},
			legend columns=4,
			enlargelimits=false,
			clip mode=individual,  %
		]
		\addplot[
			const plot,
			color=Paired-B,
			very thick,
		]
		table[col sep=tab] {figurestikz/data/ecdf_concurrentsame/dlbufferbytes-1.tsv};

		\addplot[
			const plot,
			color=Paired-F,
			very thick,
		]
		table[col sep=tab] {figurestikz/data/ecdf_concurrentsame/dlbufferbytes-2.tsv};

		\addplot[
			const plot,
			color=Paired-D,
			very thick,
		]
		table[col sep=tab] {figurestikz/data/ecdf_concurrentsame/dlbufferbytes-3.tsv};

		\addplot[
			const plot,
			color=Paired-C,
			very thick,
			dashed,
		]
		table[col sep=tab] {figurestikz/data/ecdf_concurrentsame/dlbufferbytes-4.tsv};

	\end{axis}
\end{tikzpicture}%
	}

	\subfloat[\gls{dl} throughput.]{
		\centering
\begin{tikzpicture}

	\begin{axis}[%
			width=\fwidth,
			height=0.3\fheight,
			at={(0\fwidth,0\fheight)},
			scale only axis,
			xmin=0,
			xmax=14,
			xlabel style={font=\footnotesize\color{white!15!black}},
			xlabel={Throughput [Mbps]},
			ymin=0,
			ymax=1,
			ylabel style={font=\footnotesize\color{white!15!black}},
			ylabel={ECDF},
			axis background/.style={fill=white},
			axis x line*=bottom,
			axis y line*=left,
			xmajorgrids,
			ymajorgrids,
			enlargelimits=false,
			clip mode=individual,  
		]

		\addplot[
			const plot,
			color=Paired-B,
			very thick,
		]
		table[col sep=tab] {figurestikz/data/ecdf_concurrentsame/txbrateDownlinkMbps-1.tsv};

		\addplot[
			const plot,
			color=Paired-F,
			very thick,
		]
		table[col sep=tab] {figurestikz/data/ecdf_concurrentsame/txbrateDownlinkMbps-2.tsv};

		\addplot[
			const plot,
			color=Paired-D,
			very thick,
		]
		table[col sep=tab] {figurestikz/data/ecdf_concurrentsame/txbrateDownlinkMbps-3.tsv};

		\addplot[
			const plot,
			color=Paired-C,
			very thick,
			dashed,
		]
		table[col sep=tab] {figurestikz/data/ecdf_concurrentsame/txbrateDownlinkMbps-4.tsv};

	\end{axis}
\end{tikzpicture}%
	}

	\caption{Profiles of xApps executed both individually and at the same time: both applications sending messages at $\SI{1}{\second}$ interval each, \gls{ue} with \SI{2}{\mega\bps} traffic. The dashed line represents the predicted \ran performance.}
	\label{fig:ecdf_concurrent}
\end{figure}
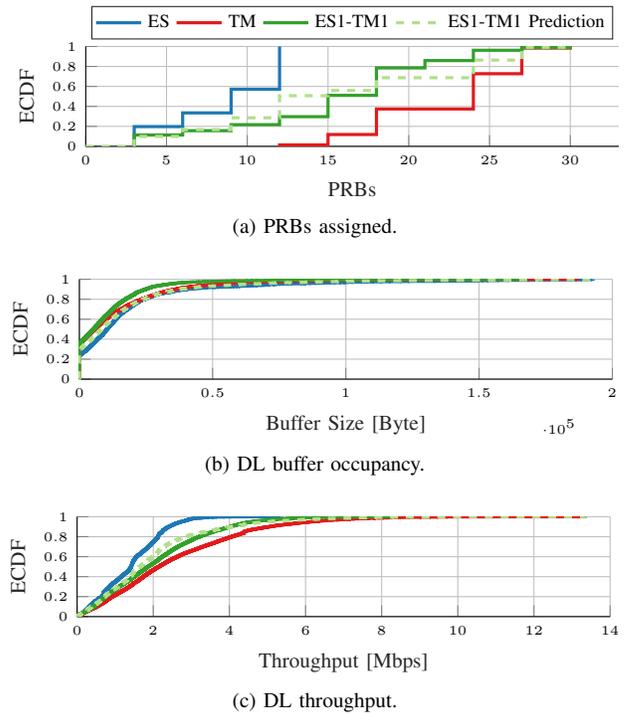

\section{RAN Performance Prediction}\label{sec:results}

In this section, we analyze the ability of our predictor to predict \ran performance based on Algorithm~\ref{alg:weighted_average} and the offline-computed profiles of the individual xApps.
We run experiments to observe how the apps perform when they are run concurrently but take decisions at different timescales.

\begin{table}[h!]
	\centering
	\caption{xApps configuration time intervals}
	\label{tab:configurations}
	\begin{tabular}{@{}lrr@{}}
		\toprule
		Configuration & $\tau_\text{\es}$ [s] & $\tau_\text{\tm}$ [s] \\
		\midrule
		ES1-TM1       & 1                     & 1                     \\
		ES2-TM10      & 2                     & 10                    \\
		ES10-TM2      & 10                    & 2                     \\
		\bottomrule
	\end{tabular}
\end{table}

We define multiple configurations for running the xApps concurrently based on the time interval $\tau$ between consecutive control actions (e.g., a new slicing policy) sent to the \gls{bs}, as shown in Table~\ref{tab:configurations}.
In the first configuration, called \textit{ES1-TM1}, both applications generate control actions at the same periodicity $\tau_\text{\es} = \tau_\text{\tm}$, while in the others, called \textit{ES2-TM10} and \textit{ES10-TM2}, one application sends a message every \SI{2}{\second} while the other sends a message every \SI{10}{\second}.

Figure~\ref{fig:ecdf_concurrent} shows the profiles for the \gls{embb} slice of the two xApps using individual offline profiling as well as when they run at the same time sending messages at the same frequency (configuration ES1-TM1), together with the predicted \gls{ecdf} represented by the dashed line.
The predicted performance and the actual performance when both applications are executed concurrently are remarkably close across all variables (also including those not depicted in the figure).

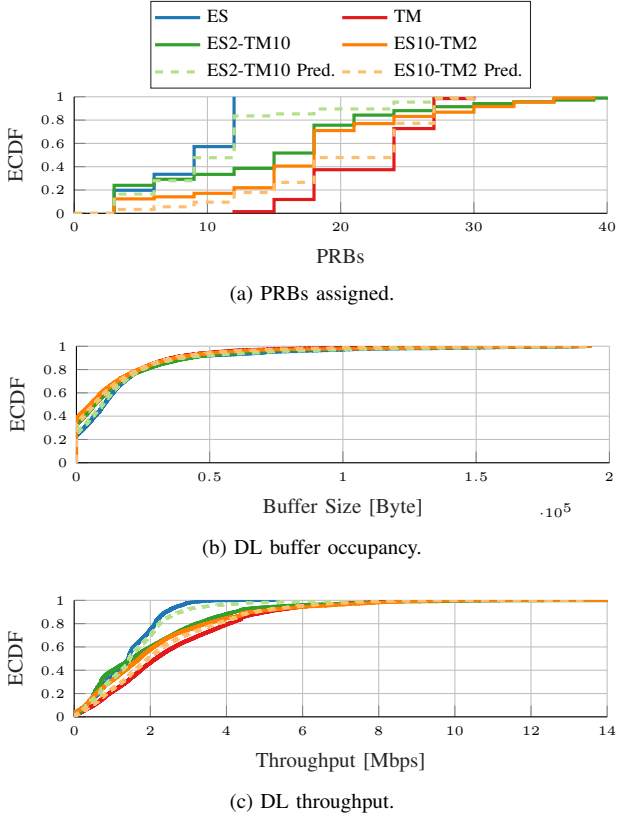
\begin{figure}[htbp]
	\centering
	\subfloat[\glspl{prb} assigned.]{
		\centering
		\begin{tikzpicture}

	\begin{axis}[%
			width=\fwidth,
			height=0.35\fheight,
			at={(0\fwidth,0\fheight)},
			scale only axis,
			xmin=0,
			xmax=40,
			xlabel style={font=\footnotesize\color{white!15!black}},
			xlabel={PRBs},
			ymin=0,
			ymax=1,
			ylabel style={font=\footnotesize\color{white!15!black}},
			ylabel={ECDF},
			axis background/.style={fill=white},
			axis x line*=bottom,
			axis y line*=left,
			xmajorgrids,
			ymajorgrids,
			legend style={
					inner sep=1pt,
					legend cell align=left,
					align=left,
					draw=white!15!black,
					font=\scriptsize,
					at={(0.5,1.07)}, anchor=south,
				},
			legend columns=2,
			enlargelimits=false,
			clip mode=individual,  %
		]
		\addplot[
			const plot,
			color=Paired-B,
			very thick,
		] table[col sep=tab] {figurestikz/data/ecdf_concurrentdiff/sliceprb-1.tsv};

		\addplot[
			const plot,
			color=Paired-F,
			very thick,
		] table[col sep=tab] {figurestikz/data/ecdf_concurrentdiff/sliceprb-2.tsv};

		\addplot[
			const plot,
			color=Paired-D,
			very thick,
		] table[col sep=tab] {figurestikz/data/ecdf_concurrentdiff/sliceprb-3.tsv};

		\addplot[
			const plot,
			color=Paired-H,
			very thick,
		]
		table[col sep=tab] {figurestikz/data/ecdf_concurrentdiff/sliceprb-4.tsv};

		\addplot[
			const plot,
			color=Paired-C,
			very thick,
			dashed,
		]
		table[col sep=tab] {figurestikz/data/ecdf_concurrentdiff/sliceprb-5.tsv};

		\addplot[
			const plot,
			color=Paired-G,
			very thick,
			dashed,
		]
		table[col sep=tab] {figurestikz/data/ecdf_concurrentdiff/sliceprb-6.tsv};

		\addlegendentry{ES}
		\addlegendentry{TM}
		\addlegendentry{ES2-TM10}
		\addlegendentry{ES10-TM2}
		\addlegendentry{ES2-TM10 Pred.}
		\addlegendentry{ES10-TM2 Pred.}

	\end{axis}
\end{tikzpicture}%
	}

	\subfloat[\gls{dl} buffer occupancy.]{
		\centering
		\begin{tikzpicture}

	\begin{axis}[%
			width=\fwidth,
			height=0.35\fheight,
			at={(0\fwidth,-0.33\fheight)},
			scale only axis,
			xmin=0,
			xmax=2e+05,
			xlabel style={font=\footnotesize\color{white!15!black}},
			xlabel={Buffer Size [Byte]},
			ymin=0,
			ymax=1,
			ylabel style={font=\footnotesize\color{white!15!black}},
			ylabel={ECDF},
			axis background/.style={fill=white},
			axis x line*=bottom,
			axis y line*=left,
			xmajorgrids,
			ymajorgrids,
			legend style={
					inner sep=1pt,
					legend cell align=left,
					align=left,
					draw=white!15!black,
					font=\scriptsize,
					at={(0.95,0.05)}, anchor=south east,
				},
			legend columns=2,
			enlargelimits=false,
			clip mode=individual,  %
		]

		\addplot[
			const plot,
			color=Paired-B,
			very thick,
		] table[col sep=tab] {figurestikz/data/ecdf_concurrentdiff/dlbufferbytes-1.tsv};

		\addplot[
			const plot,
			color=Paired-F,
			very thick,
		] table[col sep=tab] {figurestikz/data/ecdf_concurrentdiff/dlbufferbytes-2.tsv};

		\addplot[
			const plot,
			color=Paired-D,
			very thick,
		] table[col sep=tab] {figurestikz/data/ecdf_concurrentdiff/dlbufferbytes-3.tsv};

		\addplot[
			const plot,
			color=Paired-H,
			very thick,
		]
		table[col sep=tab] {figurestikz/data/ecdf_concurrentdiff/dlbufferbytes-4.tsv};

		\addplot[
			const plot,
			color=Paired-C,
			very thick,
			dashed,
		]
		table[col sep=tab] {figurestikz/data/ecdf_concurrentdiff/dlbufferbytes-5.tsv};

		\addplot[
			const plot,
			color=Paired-G,
			very thick,
			dashed,
		]
		table[col sep=tab] {figurestikz/data/ecdf_concurrentdiff/dlbufferbytes-6.tsv};

	\end{axis}
\end{tikzpicture}%
	}

	\subfloat[\gls{dl} throughput.]{
		\centering
\begin{tikzpicture}

	\begin{axis}[%
			width=\fwidth,
			height=0.35\fheight,
			at={(0\fwidth,0\fheight)},
			scale only axis,
			xmin=0,
			xmax=14,
			xlabel style={font=\footnotesize\color{white!15!black}},
			xlabel={Throughput [Mbps]},
			ymin=0,
			ymax=1,
			ylabel style={font=\footnotesize\color{white!15!black}},
			ylabel={ECDF},
			axis background/.style={fill=white},
			axis x line*=bottom,
			axis y line*=left,
			xmajorgrids,
			ymajorgrids,
			enlargelimits=false,
			clip mode=individual,  
		]

		\addplot[
			const plot,
			color=Paired-B,
			very thick,
		] table[col sep=tab] {figurestikz/data/ecdf_concurrentdiff/txbrateDownlinkMbps-1.tsv};

		\addplot[
			const plot,
			color=Paired-F,
			very thick,
		] table[col sep=tab] {figurestikz/data/ecdf_concurrentdiff/txbrateDownlinkMbps-2.tsv};

		\addplot[
			const plot,
			color=Paired-D,
			very thick,
		] table[col sep=tab] {figurestikz/data/ecdf_concurrentdiff/txbrateDownlinkMbps-3.tsv};

		\addplot[
			const plot,
			color=Paired-H,
			very thick,
		]
		table[col sep=tab] {figurestikz/data/ecdf_concurrentdiff/txbrateDownlinkMbps-4.tsv};

		\addplot[
			const plot,
			color=Paired-C,
			very thick,
			dashed,
		]
		table[col sep=tab] {figurestikz/data/ecdf_concurrentdiff/txbrateDownlinkMbps-5.tsv};

		\addplot[
			const plot,
			color=Paired-G,
			very thick,
			dashed,
		]
		table[col sep=tab] {figurestikz/data/ecdf_concurrentdiff/txbrateDownlinkMbps-6.tsv};

	\end{axis}
\end{tikzpicture}%
	}

	\caption{Profiles of xApps executing individually and concurrently: \gls{ue} with \SI{2}{\mega\bps} traffic. The dashed lines represent the predicted \ran performance.}
	\label{fig:ecdf_interval}
\end{figure}

Figure~\ref{fig:ecdf_interval} shows the predicted performance when the applications run concurrently but generate actions at different time intervals (see Table~\ref{tab:configurations}).
The predicted \glspl{ecdf} for both profiles are shown as dashed lines.
The predicted \gls{kpm} profiles closely match the measured ones, except for ES2-TM10, where the control parameter slightly deviates from the prediction.
This is attributable to the ES app's algorithmic guarantee of sufficient resources to satisfy requests, combined with its incremental control logic, which prevents the network from frequently reaching low \gls{prb} allocation values.

To quantitatively compare the real \glspl{ecdf} with the predicted \glspl{cdf}, we use the same distance functions employed for computing conflicts among applications, i.e., $\dks$ and $\dint$, shown in Table~\ref{tab:distances_predictions}.
All $\dks$ distances are below $0.29$, except for \glspl{prb} in case ES2-TM10 ($0.45$), while all $\dint$ distances are very close to $0$ (the highest being $0.12$), indicating that the predicted and real curves exhibit nearly identical trends.
For both distance metrics, \glspl{kpm} show lower distances than the control parameter. This reflects the lower \gls{int} distances observed between the two applications themselves, as their individual curves are inherently closer to each other for \glspl{kpm} than for \gls{prb} allocation.

\begin{table}[htbp]
	\centering
	\caption{Distances between experimental and predicted performance for \gls{embb} \es and \tm xApps}
	\label{tab:distances_predictions}
	\includegraphics[width=1.0\columnwidth]{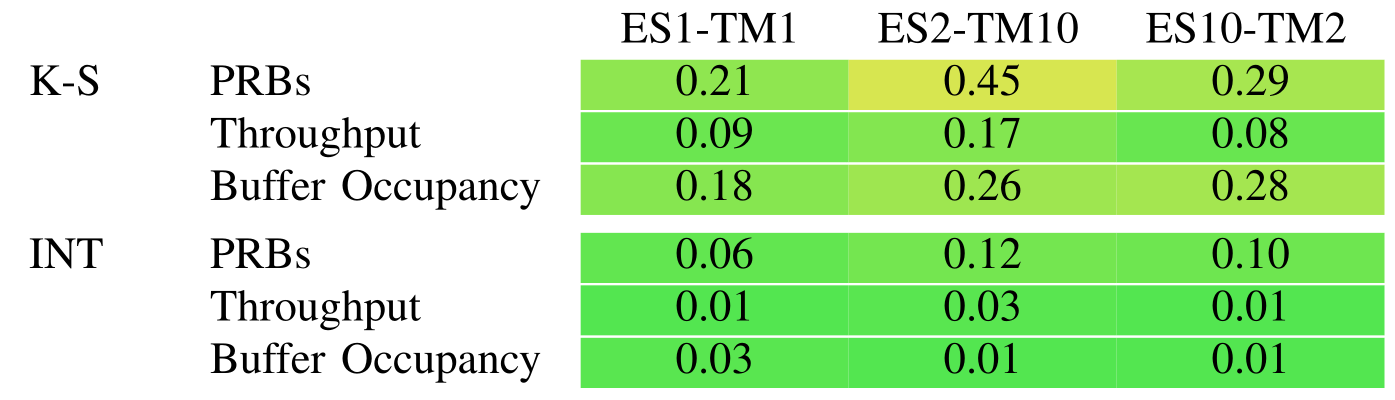}
\end{table}

\section{Conclusions}\label{sec:conclusions}

In this work, we develop a novel system to predict the impact of conflicts on the \ran for different timescale configurations.
Based on the statistical profiling of \pacifista, we use the \glspl{ecdf} and the information about the timescale configurations of concurrent applications to estimate a new \gls{cdf} for each variable that represents the predicted behavior of the \ran.
We validate this system by testing two new xApps for energy saving and throughput maximization, and show that the impact of conflicts on \ran performance can be forecast.
The proposed frequency-weighted combining relies on a last-writer-wins approximation that assumes each application's decision-making logic remains representative when co-deployed.
While results demonstrate effectiveness, accuracy may degrade under strong feedback coupling between applications.
Future work will focus on formally characterizing these applicability boundaries and extending the method to capture non-linear application interactions, and validating it across a broader set of xApp
combinations and traffic scenarios.

\bibliographystyle{IEEEtran}
\bibliography{references}

\end{document}